\documentclass[12pt]{article}
\usepackage{amsfonts}
\usepackage{color,caption}
\usepackage{graphicx}
\usepackage{amsmath}
\usepackage{float}
\usepackage{amssymb}
\providecommand{\U}[1]{\protect\rule{.1in}{.1in}}
\textwidth=6.5in \hoffset=-.75in \voffset=-1in \numberwithin{equation}{section}
\numberwithin{figure}{section}

\newcommand {\be}{\begin{equation}}
 \newcommand {\ee}{\end{equation}}
 \newcommand {\bea}{\begin{eqnarray}}
 
 \newcommand {\eea}{\end{eqnarray}}
 
\def \th {\theta}
\def \O {\Omega}
\def \f {\frac}

\def \D {\Delta}

\begin{document}
\begin{titlepage}
\bigskip \begin{flushright}
\end{flushright}
\vspace{1cm}
\begin{center}

{\Large \bf {Hidden conformal symmetry for Kerr-Newman-NUT-AdS black holes }}\\
\vskip 1cm
\end{center}
\vspace{1cm}
\begin{center}
M. F. A. R. Sakti$^{a,b,} ${ \footnote{mus561@usask.ca}}, A. M.
Ghezelbash$^{a,}${ \footnote{amg142@campus.usask.ca}}, A. Suroso$^{b,c,}${\footnote{agussuroso@fi.itb.ac.id}}, F. P. Zen$^{b,c,}${ \footnote{fpzen@fi.itb.ac.id}}
\\
$^a $Department of Physics and Engineering Physics, University of Saskatchewan,
Saskatoon, Saskatchewan S7N 5E2, Canada\\
$^b $Theoretical Physics Lab., THEPI Division, Institut Teknologi Bandung, Jl. Ganesha 10, Bandung, 40132, Indonesia \\
$^c $Indonesia Center for Theoretical and Mathematical Physics (ICTMP), Institut Teknologi Bandung, Jl. Ganesha 10 Bandung, 40132, Indonesia \\
\vspace{1cm}
\end{center}

\begin{abstract}
We show the generic non-extremal Kerr-Newman-NUT-AdS black holes are holographically dual to the hidden conformal field theories in two different pictures. The two pictures can be merged together to the dual CFTs in general picture, that are generated by the $SL(2,\mathbb{Z})$ modular group. We also extend the calculation to the extremal limit and find the corresponding quantities.
Using the extended central charges from the Kerr/CFT correspondence, we also find agreement between the macroscopic and microscopic entropies. We also find the absorption cross-section of the scalar probes for the generic and extremal Kerr-Newman-NUT-AdS black holes,
to further support the different dual CFTs to the black holes. 
\end{abstract}
\end{titlepage}\onecolumn
\bigskip

\section{Introduction}
In the last decade, the correspondence between the rotating black holes and the dual conformal field theory (CFT), was explored extensively. Initially, the correspondence was used to relate 
the physical quantities, associated to an extremal Kerr black hole, to that of a CFT \cite{stro}.  The microscopic Bekenstein-Hawking entropy, as well as the near-super radiant modes of the extremal Kerr black holes were only a few examples of black hole quantities, which could be related to a conformal symmetry. The conformal symmetry is generated by the diffeomorphisms of the near-horizon geometry of the Kerr black holes.   
Later, the correspondence was explored in vast varieties of the extremal rotating black holes, in four and higher dimensions \cite{KerrCFT1}-\cite{SaktiAnnPhys2020}, where in general, the   
near-horizon geometry includes a copy of the AdS space. The enhancement of isometries of the AdS, provides the underlying Virasoro algebra for a conformal structure. 

For the non-extremal rotating black holes, there is no known AdS structure in the near-horizon geometry of the black holes. For the non-extremal rotating black holes, we may look at the symmetries of the 
solution space of a probe field, in the background of rotating non-extremal black holes. These symmetries can be used, to find out about the possible dual CFT to the black holes \cite{cms}. The conformal invariance of the former theory, is known as the hidden conformal symmetry. The hidden CFT was explored for different types of the non-extremal rotating black holes in four and higher dimensions \cite{othe}-\cite{KSChen}.

Moreover, it was found that for certain four-dimensional rotating charged black holes, there is a possibility to find more than one dual CFT.  The four-dimensional Kerr-Newman is a case, where there are two dual CFTs to the black hole. The first CFT is dual to the rotational degree of freedom of the black hole, and the second CFT is dual to the electric charge of the black hole  \cite{Chen-JQpic}. These two different dual CFTs to the black hole, are called $J$ and $Q$ pictures, respectively  \cite{chen-SL2Z}.
The Kerr-Sen black hole is another four-dimensional rotating charged black hole \cite{sen}. Though one may naively expect there are two different CFTs dual to the black hole, however there is only one dual CFT to the Kerr-Sen black hole. In fact, in \cite{Ghezelbash}, the authors showed that there is no well defined $Q$ picture, for the Kerr-Sen black hole, in contrast to the well defined $Q$ picture for the Kerr-Newman black hole. One reason for the absence of the $Q$ picture for the Kerr-Sen geometry, is that the non-gravitational fields don’t contribute to the central charge of the dual conformal field theory \cite{GH2}.

Inspired by the existence of two CFT pictures for the four-dimensional non-extremal rotating charged Kerr-Newman black holes and the absence of the $Q$ picture for the four-dimensional non-extremal rotating charged Kerr-Sen black holes, in this article, we investigate all possible CFT pictures for the four-dimensional non-extremal rotating charged Kerr-Newman-NUT-AdS black holes. 

The presence of the NUT twist, as well as the cosmological constant make a rich structure for the horizons of the Kerr-Newman-NUT-AdS black holes. This in turn, needs a careful analysis of the wave equation of a test particle, in the background of the black holes, to reveal the conformal structure.  In fact, we consider the field equations of a charged scalar field in the background of the Kerr-Newman-NUT-AdS black holes, and search for all different possible conformal symmetries. We also find the scattering  cross-section of the scalar fields, in the near region of the non-extremal, as well as the extremal Kerr-Newman-NUT-AdS black holes. 

We organize the article as: In section \ref{gen-hidden}, we discuss in detail the properties of the Kerr-Newman-NUT-AdS black holes and consider the field equation of  a massless charged scalar field in the background of the black holes. 
In section \ref{hiddensym}, we find that the radial part of the field equation in the near region, can be written in terms of $SL(2,\mathbb{R})_L \times SL(2,\mathbb{R})_R$ Casimir operators in three different pictures.
In section \ref{exthiddensym}, we consider the extremal Kerr-Newman-NUT-AdS black holes and establish the conformal structures in three different pictures. We use a set of different conformal coordinates for the extremal black holes,  to establish the holographic pictures.
In section \ref{entropy}, we find the near horizon geometry of the Kerr-Newman-NUT-AdS black holes and calculate the central charges and the entropy of CFT in three different pictures.
In section \ref{scat},  we consider the non-extremal Kerr-Newman-NUT-AdS black holes and 
find the absorption cross section of the scalar fields. We show that the results can be obtained from the finite temperature absorption cross section in a CFT. 
In section \ref{extscat},  we consider the extremal Kerr-Newman-NUT-AdS black holes and 
find the absorption cross section of the scalar fields. Similar to non-extremal case, we show that the results can be obtained from the finite temperature absorption cross section in a CFT. 

\section{Charged scalar probe around Kerr-Newman-NUT-AdS black holes}\label{gen-hidden}

The Kerr-Newman-NUT-AdS black hole  is the exact solution to Einstein-Maxwell equations with non-vanishing NUT charge and the cosmological constant. The solutions portray the space-time for a rotating electrically and magnetically charged massive object with twisting NUT charge. The line element of Kerr-Newman-NUT-AdS space-time read as \cite{Demianski1976}-\cite{Podolsky2009} 
\begin{eqnarray}
ds^2 &=& - \frac{\Delta_r}{\Xi^2\varrho^2} \left[ dt \!-\! \{a \sin^2\theta + 2n(1-\cos\theta) \} d\phi \right]^2 +  \frac{\varrho ^2}{\Delta_r}dr^2 + \f{\varrho ^2}{\Delta_\th} d\theta^2 \nonumber\\ 
& & + \frac{\Delta_\theta \sin^2\theta}{\Xi^2 \varrho ^2}\left[adt - \{r^2+(a+n)^2 \} \right] d\phi^2,\label{KNNUTAdSmetric} \
\end{eqnarray}
where
\be
\varrho^2=r^2+(n+a\cos\th)^2 ,\label{rho}\ee
\be
\Delta_r=r^2- 2M r+e^2 +g^2 \textcolor{black}{+} \frac{r^2(r^2+6n^2+a^2)}{l^2}\textcolor{black}{+}\frac{(3n^2-l^2)(a^2-n^2)}{l^2} ,\label{Del}\ee
\be
\D_\th=1\textcolor{black}{-}\frac{a\cos\th}{l^2}(4n+a\cos\th),\, 
\Xi=1\textcolor{black}{-} \frac{a^2}{l^2}, ~ \Lambda = \textcolor{black}{-}\frac{3}{l^2}.
\label{deltainkerr}
\ee
The parameters $a$, $l$ and $n$ represents the rotation parameter, cosmological constant parameter and the NUT charge, respectively. The mass parameter, electric charge and magnetic charge of space-time are given by $M$, $e$ and $g$, respectively. The metric (\ref{KNNUTAdSmetric}), in the limit of $a=0$, reduces to the Reissner-Nordstrom-NUT-AdS black hole. Moreover, it reduces to Taub-NUT-AdS spacetime and Schwarzschild-AdS spacetime, in the limits of $a=0,~Q=0$  and $a=0,~Q=0,~n=0$, respectively. In the extremal limit of $a=0,~Q=0,~n=0,\Lambda =0$, it reduces to Schwarzschild  spacetime. The electromagnetic potential is given by  \cite{Podolsky2009}

\begin{eqnarray}
A_\mu dx^\mu &=& \frac{-er\left[a dt- \left( (a+n)^2 - (n+a \text{cos}\theta)^2 \right) d\phi \right]}{a\varrho ^2 \Xi}  \nonumber\\
& &- \frac{g(n+a\text{cos}\theta)\left[a dt- \left( r^2 + (a+n)^2 \right) d\phi \right]}{a\varrho ^2 \Xi} .\ \label{eq:electromagneticpotential}
\end{eqnarray}

The Bekenstein-Hawking entropy, Hawking temperature, angular velocity and the electric potential on the event horizon of the black hole $r_+$,
are given by
\begin{eqnarray} \label{en}
&&S_{BH}=\f{\pi}{\Xi} \left[r_+^2+(a+n)^2\right],\label{en1}\\
&&T_H=\f{2(r_+-M)l^2\textcolor{black}{+}2r_+(2r_+^2+6n^2+a^2)}{4\pi(r_+^2+(a+n)^2)l^2 \Xi},\label{en2}\\
&&\O_H=\f{a}{r_+^2+(a+n)^2},\label{en3}\\
&&\Phi_H=\f{er_+}{\left[r_+^2+(a+n)^2 \right]\Xi},\label{en4}
\end{eqnarray} 
respectively. \par
To investigate and explore the hidden conformal symmetry, we consider a massless charged scalar probe in the background of the Kerr-Newman-NUT-AdS black holes (\ref{KNNUTAdSmetric}). The non-minimally coupled field equation for the scalar probe, is given by
\begin{equation}
(\nabla_{\alpha} - i q A_{\alpha})(\nabla^{\alpha} - i q A^{\alpha})
\Phi = 0\label{KG1},
\end{equation}
where $q$ is the charge of the scalar field.  We notice the Kerr-Newman-NUT-AdS black holes (\ref{KNNUTAdSmetric}) have two translational Killing vectors $\partial_t$ and $\partial_\phi$. Hence we separate the coordinates in the scalar field as 
\cite{cms,Compere}
\begin{equation}
\Phi(t, r, \theta, \phi) = \mathrm{e}^{- i \omega t + i m \phi} R(r) S(\theta)\label{phi-expand1}.
\end{equation}
Plugging equation (\ref{phi-expand1}) in (\ref{KG1}), leads to two differential equations for the angular $S(\theta)$ and radial $R(r)$ wave functions,
\be
\frac1{\sin\theta} \partial_\theta (\sin\theta \, \partial_\theta )S(\theta) - \f{\Xi^2}{\D_\th}\left[ \frac{m^2}{\sin^2\theta}+\f{\left[2n(\cos\theta-1)-a \sin^2\theta\right]^2\omega^2}{\sin^2\theta} \right] S(\theta) \nonumber\\ \ee
\be - \f{\Xi^2}{\D_\th}\left[ 2am\omega \cos\th (4n+a\cos\th) + K_{l'} \right]S(\theta) =0  \label{angular1}, \ee
\be \partial_r (\Delta_r \partial_r)R(r) + \left[ \frac{[ (r^2 + (a+n)^2) \omega - qe  r - m a ]^2\Xi^2}{\Delta_r} + 2 a m \omega\Xi^2 - K_{l'} \right] R(r) = 0\label{radial1}, \ee
where $K_{l'}$ is the separation constant. For simplicity, we set  $ g=0 $ in the angular and radial equations (\ref{angular1}) and (\ref{radial1}).

We consider the radial equation (\ref{angular1}) at the near-horizon region, which is defined by $ \omega r \ll 1 $. Moreover, we consider the low-frequency limit for the scalar field $ \omega M \ll 1 $ and small electric charge for the scalar field $ q e \ll 1 $. Consequently, we also impose $ \omega a \ll 1, \, \omega e \ll 1, $ and $ \omega n \ll 1 $. Furthermore, in the vicinity of near-horizon region, we 
approximate the quartic $\Delta_r $ by a quadratic polynomial in  $ r-r_+ $ as,
\be
\Delta_r \simeq K(r-r_+)(r-r_*), \label{APP}
\ee 
where 
\begin{eqnarray} 
K &=& 1\textcolor{black}{+} \f{6r_+^2 +6n^2 +a^2}{l^2}, \\
r_* &=& r_+ - \f{1}{K r_+}\left[r_+^2 -e^2 \textcolor{black}{-} \f{r_+^2(r_+^2 +6n^2 +a^2)}{l^2}\textcolor{black}{-}\f{(3n^2-l^2)(a^2-n^2)}{l^2} \right]. \
\end{eqnarray}
We should notice that approximating the quartic $\Delta_r $ by a quadratic polynomial of $ r-r_+ $ is necessary to set up the radial equation, in a suitable form for exploring its hidden conformal symmetry. Such an approximation was used before for the quartic metric function of the Kerr-Newman-AdS \cite{ChenJHEP2010, ChenNuc2011-a} and Kerr-NUT-AdS \cite{ChenNuc2011-b} black holes. We also notice $ r_*  $, in general is not the inner horizon. 
Using (\ref{APP}), the radial equation (\ref{radial1}) reduces to 
\begin{equation}
\partial_r \left[(r-r_+)(r-r_*) \partial_r\right]R(r) + \left[ \f{r_+ - r_*}{r - r_+} A  \!+\! \f{r_+ - r_*}{r - r_*} B + C \right] R(r) = 0 , \label{radeq}
\end{equation} 
where
\begin{eqnarray}
&& A= \frac{ \Xi^2\left[ (r_+^2 \!+\! (a+n)^2) \omega - a m - qe r_+ \right]^2}{K^2 (r_+ - r_*)^2}, \label{A}\\ 
&& B= \frac{\Xi^2\left[ (r_*^2 \!+\! (a+n)^2) \omega - a m -qer_* \right]^2}{K^2(r_+ - r_*)^2} ,\label{B}\\
&& C= \f{q^2 e^2 -l'(l'+1)}{K} , \label{C}\
\end{eqnarray}
where we set the separation constant $K_{l'}=l'(l'+1)$.
\par
The presence of the electric charge in the scalar wave equation (\ref{KG1}) in the background of the Kerr-Newman-NUT-AdS black holes, leads to fascinating features in Kerr/CFT correspondence. The features were studied before for the Kerr-Newman as well as Kerr-Sen black holes \cite{Chen-JQpic, Ghezelbash}. In fact, we can find two different individual CFTs, which are holographically dual to the Kerr-Newman-NUT-AdS black holes,  where we call them as $J$ picture and $Q$ picture. In $J$ picture, we assume the charge of the probe is very small, so we set it equal to zero. On the other hand, in $Q$ picture, we consider the scalar probe co-rotates with the horizon,  and so it is in zero mode $ m=0 $. The two $J$ and $Q$ pictures for the Kerr-Newman-NUT-AdS black holes, are generalization of the pictures for the Kerr-Newman black holes \cite{Chen-JQpic}. 
We should note that not all charged rotating black holes are holographically dual to two different CFTs.  Specifically, the four-dimensional charged rotating Kerr-Sen black holes (which are 
the solutions to the low energy limit of heterotic string theory), do not possess the hidden conformal symmetry in a well-defined $Q$ picture \cite{Ghezelbash}.
\par
To realize the existence of two dual CFTs to the black hole, we may consider the scalar probe expansion as \cite{Chen-JQpic}
\be
\Phi = \mathrm{e}^{-i\omega t + im\phi + iq\chi} R(r) S(\theta),\label{5dwave}
\ee
where we introduce an additional internal degree of freedom $\chi$ for the scalar probe, very similar to the $U(1)$ symmetry for the coordinate $\phi$. The coordinate $\chi$ plays the role of a fibered coordinate, to uplift the four-dimensional black hole solutions to the five-dimensional solutions.
The realization of scalar field as (\ref{5dwave}), has been considered to investigate the Kerr/CFT correspondence for the other charged rotating black holes \cite{RN-CFT1}-\cite{RN-CFT4}. As we explore later, the existence of two coordinates with $U(1)$ symmetry, leads to the twofold hidden symmetry for the Kerr-Newman-NUT-AdS black holes in $J$ and $Q$ pictures. We can also combine these two different pictures into one, called the general picture, where we can apply an $SL(2,\mathbb{Z})$ modular group transformation on $\phi$ and $\chi$ coordinates.

\section{Hidden conformal symmetry for the Kerr-Newman-NUT-AdS black holes}\label{hiddensym}
In this section, we are going to investigate the hidden conformal symmetry of the Kerr-Newman-NUT-AdS black hole.   We perform the following coordinate transformations for the generic black holes 
\begin{eqnarray}
&& \omega^+ = \sqrt{\frac{r-r_+}{r-r_*}}\mathrm{e}^{2\pi T_R \phi + 2 n_R t}, \label{con1}\\
&& \omega^- = \sqrt{\frac{r-r_+}{r-r_*}}\mathrm{e}^{2\pi T_L \phi + 2 n_L t}, \label{con2}\\
&& y = \sqrt{\frac{r_+ -r_*}{r-r_*}}\mathrm{e}^{\pi (T_L +T_R) \phi + (n_L + n_R) t}. \label{conformalcoord}\
\end{eqnarray}
In terms of the new conformal coordinates $\omega^+,\,\omega^-$ and $y$, we define three locally conformal operators 
\begin{eqnarray}
&& H_1 = i \partial_+, \label{vec1}\\
&& H_{-1} = i \left(\omega^{+2}\partial_+ + \omega^{+}y\partial_y - y^2 \partial_- \right), \label{vec2}\\
&& H_0 = i \left(\omega^{+}\partial_+ + \f{1}{2}y\partial_y \right), \label{vec3}
\end{eqnarray}
as well as 
\begin{eqnarray}
&& \bar{H}_1 = i \partial_-, \label{vvec1}\\
&& \bar{H}_{-1} = i \left(\omega^{-2}\partial_- + \omega^{-}y\partial_y - y^2 \partial_+ \right), \label{vvec2}\\
&& \bar{H}_0 = i\left(\omega^{-}\partial_- + \f{1}{2}y\partial_y \right). \label{vvec3}
\end{eqnarray}

The set of operators (\ref{vec1})-(\ref{vec3}) satisfy the $ SL(2,R) $ Lie algebra
\begin{eqnarray}
\left[H_0,H_{\pm 1} \right] = \mp iH_{\pm 1}, ~~~ \left[H_{-1},H_1 \right]=-2iH_0.  
\end{eqnarray}
A similar $ SL(2,R) $ algebra exists for the set of operators (\ref{vvec1})-(\ref{vvec3}). From any of two sets of operators, we can obtain the quadratic Casimir operator as
\begin{eqnarray}
\mathcal{H}^2 = \bar{\mathcal{H}}^2 = - H_0^2 + \frac{1}{2}(H_1 H_{-1} + H_{-1} H_{1}) = \frac{1}{4}(y^2 \partial_y^2 - y\partial_y)+y^2\partial_+ \partial_-. \label{quadraticCasimir}
\end{eqnarray}
To make it easier to notice the relation between the quadratic Casimir operator (\ref{quadraticCasimir}) and the radial equation (\ref{radeq}), it is better to bring it back in terms of coordinates $ (t,r,\phi) $. In old coordinates, the quadratic Casimir operator  is given by
\begin{eqnarray}
\mathcal{H}^2 &=& (r-r_+)(r-r_*) \partial_r^2 + (2r -r_+ -r_*)\partial_r + \frac{r_+ -r_*}{r-r_*}\left(\frac{n_L-n_R}{4\pi G}\partial_\phi -\frac{T_L-T_R}{4G}\partial_t \right)^2 \nonumber\\
& &- \frac{r_+ -r_*}{r-r_+}\left(\frac{n_L+n_R}{4\pi G}\partial_\phi -\frac{T_L+T_R}{4G}\partial_t \right)^2 , \label{CAS}
\end{eqnarray}
where $ G = n_L T_R - n_R T_L $.

In $ J $ picture, we consider no electric charge $ q=0 $ for the probe field (\ref{5dwave}) in the radial equation (\ref{radeq}). We obtain the hidden conformal symmetry by comparing the radial equation (\ref{radeq}) and the Casimir operator (\ref{CAS}). We find that the radial equation (\ref{radeq}) could be rewritten  in terms of the $ SL(2,R) $ quadratic Casimir operator as
\begin{eqnarray}
\mathcal{H}^2 R(r)=\bar{\mathcal{H}}^2 R(r)= -C R(r), \label{quadraticCasimirEq}
\end{eqnarray}
where we should identify the constants
\be
 n_L^J = -\frac{K}{2(r_+ + r_*)\Xi},~~~ n_R^J =0, \label{nJ}
\ee
\be
T_L^J = \frac{K[r_+^2 + r_*^2+2(a+n)^2]}{4\pi a(r_+ + r_*)\Xi},~~~ T_R^J =\frac{K(r_+ - r_*)}{4\pi a \Xi}. \label{tempCFTKNNUT}
\ee

In $ Q $ picture, we consider the scalar probe field with only the zero-mode of the angular momentum. So, we set $ m=0$ in the scalar probe expansion (\ref{5dwave}) as well as the radial equation (\ref{radeq}). It leads to the second holographic description of the Kerr-Newman-NUT-AdS back holes. 
In  $ Q $ picture, the radial equation (\ref{radeq}) can be rewritten as
\begin{eqnarray}
\mathcal{H}^2 R(r)=\bar{\mathcal{H}}^2 R(r)= -C R(r), \label{quadraticCasimirEq2}
\end{eqnarray}
where we identify 
\be
 n_L^Q = -\frac{K(r_+ + r_*)}{4[r_+ r_*-(a+n)^2]\Xi},~~~ n_R^Q =-\frac{K(r_+ - r_*)}{4[r_+ r_*-(a+n)^2]\Xi}, \\\label{nQ}
\ee
\be
T_L^Q = \frac{K[r_+^2 + r_*^2+2(a+n)^2]}{4\pi e[r_+ r_*-(a+n)^2]\Xi},~~~ T_R^Q =\frac{K(r_+ - r_*)}{4\pi e[r_+ r_*-(a+n)^2] \Xi}. \label{tempCFTKNNUTQ}
\ee
We note that the radial equations (\ref{quadraticCasimirEq}) and (\ref{quadraticCasimirEq2}) in $J$ and $Q$ pictures, look similar to each other, though we have two very different sets of constants (\ref{nJ}), (\ref{tempCFTKNNUT}) in $J$ picture and (\ref{nQ}),(\ref{tempCFTKNNUTQ}) in $Q $ picture, respectively.
We can combine the two underlying $U(1) $ symmetries in $J$ and $Q$ pictures, generated by the Killing vectors  $ \partial_\phi $ and $ \partial_\chi $ respectively, into a third picture. We call the third picture  as the general picture, in which there is a modular group $ SL(2,\mathbf{Z}) $ acting on the torus $ (\phi,\chi) $. The modular group $ SL(2,\mathbf{Z}) $ of the torus is given by \cite{chen-novelsl2z}-\cite{Ghezelbashdeform}
\begin{eqnarray}
\left(\begin{array}{c}
\phi' \\
\chi' \
\end{array} \right) = \left(\begin{array}{cc}
\alpha & \beta \\
\eta & \tau \
\end{array} \right)\left(\begin{array}{c}
\phi \\
\chi \
\end{array} \right), \label{modulargroup}
\end{eqnarray}
where $ \alpha, \beta, \eta, \tau $ parametrize the $ SL(2,\mathbf{Z}) $ group elements.  The modular group $ SL(2,\mathbf{Z})  $ may hint to the possible existence of a super conformal field theory, with a global $U(1)$ symmetry, in which the $J$ and $Q$ pictures are related to the spectral flow transformations. We leave this interesting subject for a future research. After performing this modular group transformation, one can find
\be
m = \alpha m' + \eta q'~~~,~~~q = \beta m' + \tau q'.\label{em-em}
\ee
The radial equation is given by 
\begin{eqnarray}
\partial_r \left(\Delta \partial_r\right)R(r) + \left[ \f{r_+ - r_*}{r - r_+} A  \!+\! \f{r_+ - r_*}{r - r_*} B + C \right] R(r) = 0,\label{radialG}
\end{eqnarray}  
where
\begin{eqnarray}
&& A= \frac{ \Xi^2 2\left[ (r_+^2 \!+\! (a+n)^2) \omega - (a\alpha +e\beta r_+)m'- (a\eta +e\tau r_+)q' \right]^2}{K^2 (r_+ - r_*)^2}, \label{AG}\\
&& B= \frac{\Xi^2\left[ (r_*^2 \!+\! (a+n)^2) \omega - (a\alpha +e\beta r_*)m'-(a\eta +e\tau r_*)q' \right]^2}{K^2(r_+ - r_*)^2} , \label{BG}\\
&& C= \f{(\beta m' + \tau q')^2 e^2 -l'(l'+1)}{K}. \label{CG}
\end{eqnarray}

In the general picture, we may choose to set $ q'=0 $ or $ m'=0 $, that we call  $ J'$ and $Q' $ pictures, respectively.  In $J'$ picture, the radial equation (\ref{radialG}) matched to the expression (\ref{CAS}) for the quadratic Casimir operator,  by choosing 
\be
 n_L^G = \frac{-K[2a\alpha +e\beta(r_+ + r_*)]}{4[a\alpha (r_+ + r_*) +e\beta (r_+ r_*-(a+n)^2)]\Xi}, \label{nG1}
\ee
\be 
n_R^G =\frac{-Ke\beta(r_+ - r_*)}{4[a\alpha (r_+ + r_*) +e\beta (r_+ r_*-(a+n)^2)]\Xi}, \label{nG2}
\ee
\be
T_L^G = \frac{K[r_+^2 + r_*^2+2(a+n)^2]}{4\pi[a\alpha (r_+ + r_*) +e\beta (r_+ r_*-(a+n)^2)]\Xi}, \label{nG3}
\ee
\be T_R^G =\frac{K(r_+^2 - r_*^2)}{4\pi[a\alpha (r_+ + r_*) +e\beta (r_+ r_*-(a+n)^2)]\Xi}. \label{tempCFTKNNUTG}
\ee

In $ Q' $ picture, we set $ q'=0 $ and the radial equation becomes similar to (\ref{radialG}) with the replacements $ (\alpha, \beta, m') $ to $ (\eta, \tau, q') $ in (\ref{radialG})  and also in the equations (\ref{AG})-(\ref{tempCFTKNNUTG}). 
We note that the equations (\ref{nJ}) and (\ref{tempCFTKNNUT}) in the $ J $ picture, can be obtained simply from (\ref{nG1})-(\ref{tempCFTKNNUTG}) by setting  $ \alpha=1, \beta=0 $. 
On the other hand, the equations (\ref{nQ}) and (\ref{tempCFTKNNUTQ}) in the $ Q $ picture, can be obtained from (\ref{nG1})-(\ref{tempCFTKNNUTG}) by setting  $ \alpha=0, \beta=1 $.
So the results for $n$'s and temperatures in the general picture include the corresponding results in both $J$ and $Q$ pictures.
We also note that our results reduce to those of \cite{ChenJHEP2010,ChenNuc2011-a}, if we set the  NUT charge to be zero. Moreover, if we set $ e=0 $, our results reduce to the results of the paper \cite{ChenNuc2011-b}.

\section{Hidden conformal symmetry for the extremal Kerr-Newman-NUT-AdS black holes}\label{exthiddensym}
In this section, we consider the extremal  Kerr-Newman-NUT-AdS black holes. In the extremal case, the conformal coordinates require to be different from the conformal coordinates for the generic black holes. In the background of the extremal Kerr-Newman-NUT-AdS black holes, the radial equation for the scalar probe becomes
\begin{equation}
\partial_r \left(\Delta \partial_r\right)R(r) + \left(  A_1  \!+\! B_1  +C \right) R(r) = 0 ,
\end{equation}  \label{extradialKNNUTAdS}
where
\begin{eqnarray}
&& A_1= \frac{ 2(2r_+ \omega - qe) \left[ (r_+^2 \!+\! (a+n)^2) \omega - a m - qe r_+ \right]\Xi^2}{K^2 (r - r_+)^2}, \label{Aex}\\
&& B_1= \frac{\left[ (r_+^2 \!+\! (a+n)^2) \omega - a m -qer_+ \right]^2 \Xi^2}{K^2(r - r_+)^2} ,\label{Bext}\\
&&\Delta = (r - r_+)^2 ,\label{Delext}
\end{eqnarray}
and $C$ is given by equation (\ref{C}). For the extremal black holes,  we implement the following conformal coordinates 
\begin{eqnarray}
&& \omega^+ = \frac{1}{2}\left(\alpha_1 t +\beta_1 \phi - \frac{\gamma_1}{r-r_+}\right),\label{extconformalcoord1}\\
&& \omega^- = \frac{1}{2}\left(\mathrm{e}^{2\pi T_L \phi + 2 n_L t}- \frac{2}{\gamma_1}\right), \\
&& y = \sqrt{\frac{\gamma_1}{2(r-r_+)}}\mathrm{e}^{\pi T_L \phi + n_Lt},\label{extconformalcoord}\
\end{eqnarray}
where $\alpha_1,\, \beta_1$ and $ \gamma_1 $ are constants. 
From the coordinates transformations (\ref{extconformalcoord1})-(\ref{extconformalcoord}), we define the two sets of locally conformal operators as (\ref{vec1})-(\ref{vvec3}) which produce an  $ SL(2,R)\times SL(2,R) $ symmetry. The quadratic Casimir operators for the symmetry algebra (\ref{quadraticCasimir}) are given by
\begin{eqnarray}
&&\mathcal{H}^2 =\bar {\mathcal{H}}^2 = \partial_r(\Delta \partial_r) - \left(\frac{\gamma_1(2\pi T_L\partial_t -2n_L \partial_\phi)}{\hat{A}(r-r_+)} \right)^2 - \frac{2\gamma_1(2\pi T_L\partial_t -2n_L \partial_\phi)}{\hat{A}^2(r-r_+)}(\beta_1 \partial_t - \alpha_1 \partial_\phi) , \nonumber\\
&&\label{CASEXT}
\end{eqnarray}
in terms of coordinates $ (t,r,\phi) $, where $ \hat A = 2\pi T_L \alpha_1 -2n_L \beta_1 $.
In $J$ picture, where $ q=0$, we find that the radial equation could be rewritten in terms of $ SL(2,R)\times SL(2,R) $ quadratic Casimir operators as
\begin{eqnarray}
\mathcal{H}^2 R(r)=\bar{\mathcal{H}}^2 R(r)= -C R(r), \label{extquadraticCasimirEq}
\end{eqnarray}
by choosing 
\be
\hspace*{14mm}\beta_1^J = \frac{\gamma_1 K}{a\Xi},~~~ \alpha_1^J = 0,\
\ee
\be
\hspace*{19mm}T_R^J = 0,~~~n_R^J =0, \label{exttempCFTKNNUTR}
\ee
\be
T_L^J = \frac{K[r_+^2 +(a+n)^2]}{4\pi a r_+\Xi},~~~ n_L^J =\frac{-K}{4\pi r_+ \Xi}. \label{exttempCFTKNNUT}
\ee
We notice the temperatures and $n$'s in (\ref{exttempCFTKNNUTR}) and (\ref{exttempCFTKNNUT}) are consistent with the temperatures and $n$'s in (\ref{nJ}) and (\ref{tempCFTKNNUT}), in the extremal limit. 

In $Q$ picture, we consider only the zero-mode of the angular momentum for the charged probe. We find that the radial equation could be rewritten in terms of $ SL(2,R)\times SL(2,R) $ quadratic Casimir operators as (\ref{extquadraticCasimirEq}), by choosing
\be
 \beta_1^Q = \frac{2\gamma_1 Kr_+}{e[\textcolor{black}{r_+^2}-(a+n)^2]\Xi},~~~ \alpha_1^Q =\frac{-K\gamma_1}{[\textcolor{black}{r_+^2}-(a+n)^2]\Xi}, \
\ee
\be
T_R^Q = 0,~~~ n_R^Q =0, \label{exttempCFTKNNUTRQ}
\ee
\be
T_L^Q = \frac{K[r_+^2 +(a+n)^2]}{2\pi e[\textcolor{black}{r_+^2}-(a+n)^2]\Xi},~~~ n_L^Q =\frac{-Kr_+}{2e[\textcolor{black}{r_+^2}-(a+n)^2] \Xi}. \label{exttempCFTKNNUTRQQ}
\ee
We notice the temperatures and $n$'s in (\ref{exttempCFTKNNUTRQ}) and (\ref{exttempCFTKNNUTRQQ}) are in agreement  with the temperatures and $n$'s in (\ref{nQ}) and (\ref{tempCFTKNNUTQ}), in the extremal limit. 

In general $ J'$ picture, we find that the radial equation could be rewritten in terms of $ SL(2,R)\times SL(2,R) $ quadratic Casimir operators as (\ref{extquadraticCasimirEq}), by choosing
\be
\beta_1 ^G = \frac{2\gamma_1 r_+ K}{[2a\alpha r_+ +e\beta (r_+^2 -(a+n)^2)]\Xi},~~~ \alpha_1^G =\frac{-\gamma_1 e\beta K}{[2a\alpha r_+ +e\beta (r_+^2-(a+n)^2)]}, \label{alphaJ'}
\ee
\be
\hspace*{-10mm}T_R^G = 0,~~~ n_R^G =0, \label{exttempCFTKNNUTRJ'}
\ee
\be
T_L^G = \frac{K[r_+^2+(a+n)^2]}{2\pi[2a\alpha r_+ +e\beta (r_+^2-(a+n)^2)]\Xi},~~~ n_L^G = \frac{-K\left(a\alpha +e\beta r_+\right)}{2[2a\alpha r_+ +e\beta (r_+^2-(a+n)^2)]\Xi}. \label{exttempCFTKNNUTLJ'}
\ee
We note the temperatures and $n$'s in (\ref{exttempCFTKNNUTRJ'}) and (\ref{exttempCFTKNNUTLJ'}) are in agreement  with the temperatures and $n$'s in equations (\ref{nG1})-(\ref{tempCFTKNNUTG}), in the extremal limit. 
Moreover, in $Q'$ picture, we find $\alpha_1,\beta_1$, temperatures and $n$'s are given by expressions (\ref{alphaJ'})-(\ref{exttempCFTKNNUTLJ'}) with replacing $ (\alpha, \beta, m') $ to $ (\eta, \tau, q') $, respectively.

\section{Microscopic entropy description}\label{entropy}

The existence of the hidden conformal symmetry suggests that the Kerr-Newman-NUT-AdS black holes could be portrayed by the dual CFTs.  For the generic black holes, the dual CFTs have non-zero left- and right-moving temperatures.  For the  extremal black holes,  the CFTs have only non-zero left temperature.  The other essential information for any CFT are the central charges of the theory. For the extremal black holes, the central charges of the dual CFT  
could be derived from an analysis of the asymptotic symmetry group \cite{stro,Compere}. 

To find the near-horizon geometry of the extremal  Kerr-Newman-NUT-AdS black holes (\ref{KNNUTAdSmetric}),  
we consider the following coordinate transformations
\begin{eqnarray}
r = r_++\lambda r_0 y,~~ t = \frac{r_0 \Xi}{\lambda}\tau, ~~ \phi = \varphi + \frac{\Omega_H r_0 \Xi}{\lambda}\tau, \label{extremaltransformation} \
\end{eqnarray}
where $r_0=\sqrt{{r_+^2+(a+n)^2}}$ and $ \lambda \rightarrow 0 $ shows the near-horizon limit.
The metric (\ref{KNNUTAdSmetric}) becomes
\begin{eqnarray}
ds^2 = \Gamma(\theta)\left(-\f{K^2 y^2}{\Xi^2} d\tau^2 + \frac{dy^2}{y^2} + \alpha(\theta) d\theta ^2 \right) +\gamma(\theta) \left(d\varphi +py d\tau\right)^2, \label{extremalmetric}\
\end{eqnarray}
where
\begin{equation}
\Gamma(\theta)=\frac{\varrho_+ ^2}{K}, ~~\alpha(\theta) = \frac{K}{\Delta_\theta}, ~~\gamma(\theta)=\frac{r_0^4 \Delta_\theta \sin^2\theta}{\varrho_+^2 \Xi^2}, \label{GAM}
\end{equation}
and
\begin{equation}
\varrho_+^2 = r_+^2 + (n + a\cos\theta)^2, ~~p = \frac{2ar_+}{r_0^2}. \label{GAM1}\
\end{equation}
Using the scaling $ d\tau \rightarrow \Xi d\tau/K $, we obtain
\begin{eqnarray}
ds^2 = \Gamma(\theta)\left(-y^2 d\tau^2 + \frac{dy^2}{y^2} + \alpha(\theta) d\theta ^2 \right) +\gamma(\theta) \left(d\varphi +\hat{p}y d\tau\right)^2, \label{extremalmetric}\
\end{eqnarray}
where
\begin{equation}
\hat{p} = \frac{2ar_+\Xi}{K r_0^2}. \
\end{equation}
In the near-horizon limit, we find that the gauge field (\ref{eq:electromagneticpotential}) is given by
\begin{eqnarray}
A_\mu dx^\mu = f(\theta) \left(d\varphi +\hat{p}yd\tau \right)-\frac{e[r_+^2-(a+n)^2]}{[r_+^2+(a+n)^2]\Xi}d\varphi ,\ \label{eq:nearhorizonelectromagneticpot1}
\end{eqnarray}
where
\begin{eqnarray}
f(\theta) =\frac{er_0^2\left[(r_+^2-n^2)-a(2n+a\text{cos}\,\theta)\text{cos}\,\theta \right]}{2ar_+ \rho_+ ^2\Xi}. \label{eq:ftheta} \
\end{eqnarray}
The left-moving central charge can be read off from the near-horizon geometry of the extremal black holes. In $ J $ picture, the central charge of the CFT is given by
\begin{equation}
c_L = 3\hat{p}\int^\pi_0 d\theta\sqrt{\Gamma(\theta) \alpha(\theta)\gamma(\theta)} = \frac{12ar_+}{K}. \label{c-calc}
\end{equation}

We also propose the following left- and right-moving central charges for the CFT dual to the generic black holes as
\begin{equation}
c_L^J = c_R^J = \f{6a(r_+ +r_*)}{K}. \label{centralchargeJpic}\
\end{equation}
We notice that the central charges (\ref{centralchargeJpic}) reduce to those for dual CFT for the Kerr black holes \cite{stro, cms}, the Kerr-Newman-AdS and  the Kerr-NUT-AdS black holes \cite{ChenJHEP2010}-\cite{ChenNuc2011-b} in the proper limits of parameters.  Our proposal for the central charges (\ref{centralchargeJpic}) together with the CFT temperatures (\ref{tempCFTKNNUT}) 
lead to the Cardy entropy for the dual CFT to the Kerr-Newman-NUT-AdS black holes, as
\be
S_{CFT}=\frac{\pi^2}{3}(c_L^JT_L^J+c_R^JT_R^J)=\frac{\pi}{\Xi}[r_+^2+(a+n)^2],
\ee
in perfect agreement with the Bekenstein-Hawking entropy for the  Kerr-Newman-NUT-AdS black holes
\begin{eqnarray}
S_{BH}=\frac{\pi}{\Xi}[r_+^2+(a+n)^2].
\end{eqnarray}
In $ Q $ picture, we enhance the geometry of the black hole with a fibered coordinate that will uplift the geometry to the five-dimensional solutions \cite{RN-CFT1}-\cite{RN-CFT3}. In addition, the gauge field can be a part of the geometry, such that  we recover the near-horizon metric of the five-dimensional extremal black holes  
Applying the gauge transformations, we find the gauge bundle
\begin{eqnarray}
\textbf{A}=\hat{p}' y d\tau -\frac{en\cos\theta}{r_+}d\varphi, ~~~\hat{p}'=\frac{e(r_+^2-n^2)}{(r_+^2+n^2)}. \label{gaugeQpic}
\end{eqnarray}
We can combine the four-dimensional near horizon metric (\ref{extremalmetric}) and the gauge field  (\ref{gaugeQpic}), to write the five-dimensional near horizon metric as 
\begin{eqnarray}
ds_5^2=ds^2+(d\hat{y}+\textbf{A})^2,\label{5}
\end{eqnarray}
where $ \hat{y} $ is the fibered coordinate, with period $ 2\pi R_n $ and $ R_n $ is an integer. We also consider the boundary conditions for the deviation of the five-dimensional full metric from the near horizon metic (\ref{5}) as
\begin{eqnarray}
h_{\mu \nu} \sim \left(\begin{array}{ccccc}
\mathcal{O}(y^2) & \mathcal{O}\left(\frac{1}{y^2}\right) &  \mathcal{O}\left(\frac{1}{y}\right) & \mathcal{O}(y) &  \mathcal{O}(1) \\
 &  \mathcal{O}\left(\frac{1}{y^3}\right) &  \mathcal{O}\left(\frac{1}{y^2}\right) & \mathcal{O}\left(\frac{1}{y}\right) &  \mathcal{O}\left(\frac{1}{y}\right) \\
 &  & \mathcal{O}\left(\frac{1}{y}\right) & \mathcal{O}(1) &  \mathcal{O}\left(\frac{1}{y}\right)\\
 &  &  & \mathcal{O}\left(\frac{1}{y}\right) & \mathcal{O}(1)\\
 &  &  &  & \mathcal{O}(1)\
\end{array} \right), \label{eq:gdeviation2}
\end{eqnarray}
in the basis $ (\tau,y,\theta, \varphi, \hat{y}) $, similar to Ref. \cite{RN-CFT1}-\cite{RN-CFT3} . We find the diffeomorphisms  that preserve the boundary conditions (\ref{eq:gdeviation2}), are generated by
\begin{eqnarray}
\zeta_{\epsilon} = \epsilon(\hat{y}) \partial_{\hat{y}} - r \epsilon '(\hat{y})\partial _{{y}} ,\
\end{eqnarray}
where $ \epsilon(\hat{y})=\mathrm{e}^{-i\hat{n}\hat{y}} $. 
Using eq. (\ref{c-calc}), we find the central charge for the dual CFT to the near-horizon geometry (\ref{5})
\begin{equation}
c_L= \f{6er_+^2}{K}.\label{c5}
\end{equation}

We then propose the
generalization of the central charge (\ref{c5}) in $ Q $ picture, for the generic black holes, is given by
\begin{eqnarray}
c_L^Q=c_R^Q = \frac{6e[r_+r_* - (a+n)^2]}{K}. \label{centralchargeQpic}
\end{eqnarray}
Using the central charges (\ref{centralchargeQpic}) for the dual CFT in $Q$ picture, and the temperatures in (\ref{exttempCFTKNNUTRQ}) and (\ref{exttempCFTKNNUTRQQ}), we find, from the
Cardy formula, 
\be
S_{CFT}=\frac{\pi^2}{3}(c_L^QT_L^Q+c_R^QT_R^Q),
\ee
that 
\begin{eqnarray}
S_{CFT}=\frac{\pi}{\Xi}[r_+^2+(a+n)^2].\label{SQ}
\end{eqnarray}
The Cardy entropy (\ref{SQ}) in $Q$ picture, is obviously in agreement with the Bekenstein-Hawking entropy (\ref{en1}) for the black holes.
Moreover, the central charges in the general picture 
\begin{eqnarray}
c_L^G=c_R^G=\frac{6\left[a\alpha (r_+ +r_*)+e\beta(r_+ r_* -(a+n)^2)\right]}{K},
\end{eqnarray}
together with the temperatures in (\ref{exttempCFTKNNUTRJ'}) and (\ref{exttempCFTKNNUTLJ'}), yields the Cardy entropy which is similarly given by (\ref{SQ}), in perfect agreement with the Bekenstein-Hawking entropy (\ref{en1}).

\section{Scattering cross-section on non-extremal background}\label{scat}
To further support the correspondence between the Kerr-Newman-NUT-AdS black holes and the CFTs,  we consider the absorption cross-section of scalar probes in the background of the generic black holes.
For the near horizon region, where we can expand the metric function $ \Delta_r $  by the quadratic order in $ (r-r_+) $ as (\ref{APP}), we use the radial wave equation (\ref{radeq}) to find the scattering cross-section.  So, for a near-extremal black hole, 
we consider the following near-extremal coordinate transformations from black hole coordinates $(r,t,\phi)$ to $(y,\tau,\varphi)$, as
\begin{eqnarray}
r = \f{r_+ + r_*}{2}+\lambda r_0 y, ~~r_+ -r_* = \mu \lambda r_0,~~ t = \frac{r_0 \Xi}{\lambda}\tau, ~~ \phi = \varphi + \frac{\Omega_H r_0 \Xi}{\lambda}\tau, \label{nearextremaltransformation} \
\end{eqnarray}
where $r_0=\sqrt{{r_+^2+(a+n)^2}}$ and $ \lambda \rightarrow 0 $ shows the near-horizon limit and $\mu$ is the near-extremality parameter.
We also consider the scalar probe with frequencies around  the super-radiant bound $ \omega_s = m\Omega_H + q\Phi_H$
\begin{equation}
\omega = \omega_s +\hat{\omega}\frac{\epsilon}{r_0},
\end{equation}
where $\Omega_H$ and $\Phi_H$ are given by (\ref{en3}) and (\ref{en4}), respectively and $\epsilon \rightarrow 0$.
We can re-write the radial equation (\ref{radeq}) by
\begin{eqnarray}
\left[\partial_y\left(y-\f{\mu}{2}\right)\left(y+\f{\mu}{2}\right)\partial_y +\f{A_s}{y-\f{\mu}{2}}+\f{B_s}{y+\f{\mu}{2}}           +C_s\right]R(y) =0,\label{radialequationnearext}
\end{eqnarray}
where
\begin{eqnarray}
A_s = \frac{\hat{\omega}^2\Xi^2}{\mu}, ~~~ B_s = -\mu\Xi^2 \left(\frac{\hat{\omega}}{\mu}-\frac{2\left(m\Omega_H + q\Phi_H\right)r_+}{K} +\frac{qe}{K} \right)^2, \
\end{eqnarray}
and $ C_s $ is the separation constant. Moreover, we change the coordinate from $y$  to $z$, by 
$ z =\frac{y-\mu/2}{y+\mu/2} $, where the radial equation (\ref{radialequationnearext}) becomes
\begin{eqnarray}
\left[z(1-z)\partial_z^2 + (1-z)\partial_z +\f{\hat{A_s}}{z}+\hat{B_s}+\f{C_s}{1-z}\right]R(z) =0,\label{radialequationnearext0}
\end{eqnarray}
where
\begin{eqnarray}
\hat{A_s} = \frac{\hat{\omega}^2\Xi^2}{\mu ^2}, ~~~\hat{B_s} = -\Xi^2 \left(\frac{\hat{\omega}}{\mu}-\frac{2\left(m\Omega_H + q\Phi_H\right)r_+}{K} +\frac{qe}{K} \right)^2. 
\end{eqnarray}
The solutions to differential equation (\ref{radialequationnearext0}) are given by
\begin{equation}
R(z)= z^{\alpha}(1-z)^{\beta}F(a_s,b_s,c_s;z), \label{sol}
\end{equation}
where $ F(a_s,b_s,c_s;z) $ is the hypergeometric function with the parameters 
\begin{equation}
a_s = \beta +i(\gamma-\alpha), ~b_s =\beta -i(\gamma+\alpha),~c_s = 1- 2i\alpha, 
\end{equation}
and
\begin{equation}
\alpha = \sqrt{\hat{A_s}}, ~\beta = \frac{1}{2}\left(1-\sqrt{1-4C_s}\right),~\gamma = \sqrt{-\hat{B}_s}. \nonumber\
\end{equation}

For large values of the radial coordinate $r$ (or equivalently $y>>\mu/2 $), where  $z \sim 1$, the solutions (\ref{sol}) reduce to
\begin{eqnarray}
R(y) \sim D_1 y^{-\beta}+D_2 y^{\beta-1},
\end{eqnarray}
where
\begin{equation}
D_1 = \frac{\Gamma(c_s)\Gamma(2h-1)}{\Gamma(c_s-a_s)\Gamma(c_s-b_S)}, ~~~ D_2 = \frac{\Gamma(c_s)\Gamma(1-2h)}{\Gamma(a_s)\Gamma(b_s)}, ~~~ 
\end{equation}
and 
\begin{equation}
h= 1-\beta=\frac{1}{2}\left(1+\sqrt{1-4C_s}\right),\label{hh}
\end{equation}
is the conformal weight.
Hence, we find the absorption cross-section of the scalar fields as 
\begin{align}
P_{abs} \sim \left| D_1 \right|^{-2} = \frac{\sinh \left( {2\pi\alpha } \right)}{2\pi \alpha}\frac{{\left| {\Gamma \left( c_s - a_s  \right)} \right|^2 \left| {\Gamma \left( c_s - b_s \right)} \right|^2 }}{{ \left( {\Gamma \left( {2h-1} \right)} \right)^2 }}\label{PabsSL2Z}.
\end{align}

Alternatively, we can find the real-time correlation function $G_R $, considering $ D_1 $ as the source and $ D_2 $ as the response, by
\begin{eqnarray}
G_R \sim \frac{D_2}{D_1} = \frac{\Gamma(1-2h)}{\Gamma(2h-1)} \frac{\Gamma(c_s-a_s)\Gamma(c_s-b_S)}{\Gamma(a_s)\Gamma(b_s)},
\end{eqnarray}
such that $P_{abs}  \sim \Im(G_R)$.
To further support the correspondence between the rotating black hole (\ref{KNNUTAdSmetric}) and the two-dimensional CFT, we show that the absorption cross-section for the scalar fields  (\ref{PabsSL2Z}) can be obtained from the absorption cross-section in a two-dimensional dual CFT in three different pictures. In a two-dimensional CFT with conformal weights $h_{L,R}$ and temperatures $T_{L,R}$ , the absorption cross-section for the scalar fields with frequencies $\tilde \omega_{L,R}$, is given by \cite{Malda-Strom}
\be
P_{abs} \sim {T _L}^{2h_L - 1} {T _R}^{2h_R - 1} \sinh \left( {\frac{{{\tilde{\omega}} _L }}{{2{T _L} }} + \frac{{{\tilde{\omega}} _R }}{{2{T _R} }}} \right)\left| {\Gamma \left( {h_L + i\frac{{{\tilde{\omega}} _L }}{{2\pi {T _L} }}} \right)} \right|^2 \left| {\Gamma \left( {h_R + i\frac{{{\tilde{\omega}} _R }}{{2\pi {T _R} }}} \right)} \right|^2.\label{Pabs2CFTSL2Z}
\ee

To find agreement between (\ref{PabsSL2Z}) and (\ref{Pabs2CFTSL2Z}), we should choose proper left and right frequencies ${\tilde{\omega}} _L,{\tilde{\omega}} _R$. In order to do so, we consider the first law of thermodynamics for the general charged rotating black holes (\ref{KNNUTAdSmetric}), that is given by
\be \label{BHthermoLaw}
T_H \delta S_{BH} = \delta M - \Omega _H \delta J - \Phi _H \delta Q,
\ee 
where $T_H$, $\Omega_H$ and $\Phi_H$ are given by (\ref{en2}), (\ref{en3}) and (\ref{en4}). 
By varying Cardy entropy, we obtain
\be \label{delSCFT}
\delta S_{CFT} = \frac{{\delta E_L }}{{T_L }} + \frac{{\delta E_R }}{{T_R }}.
\ee 
Identifying $\delta M$ as $\omega$, $\delta J$ as $m$ and $\delta Q$ as $q$, yield  to  identification of $\delta E _{R,L}$ as $\tilde{\omega} _{R,L}$. Equating the 
variations of entropy in (\ref{BHthermoLaw}) and (\ref{delSCFT}) in general picture, leads to a set of left and right frequencies
\be 
\tilde \omega _{L,R} = \omega _{L,R} - q_{L,R} \mu _{L,R} ,\label{ff}
\ee
where
\begin{eqnarray}
&&\omega _{L} = \frac{(r_+ + r_*)[r_+^2 + r_*^2 +2(a+n)^2]}{2[a\alpha (r_+ + r_*)+e\beta (r_+r_* -(a+n)^2)]}\omega, \label{oL}\ \\
&&\mu _{L} = \frac{e[r_+^2 + r_*^2 +2(a+n)^2]}{2[a\alpha (r_+ + r_*)+e\beta (r_+r_* -(a+n)^2)]\Xi},~~~q_L = q, \label{muL} \\
&&\omega _{R} = \omega _{L}-\frac{2a(r_+ + r_*)\Xi }{2[a\alpha (r_+ + r_*)+e\beta (r_+r_* -(a+n)^2)]}m,  \label{oR} \\
&&\mu _{R} = \frac{e(r_+ + r_*)^2}{2[a\alpha (r_+ + r_*)+e\beta (r_+r_* -(a+n)^2)]}, ~~~q_R = q.\label{muR} \
\end{eqnarray}

We note that the conformal weights $ h_{L,R} $ in (\ref{Pabs2CFTSL2Z}) are equal to  $h $, as in (\ref{hh}). We also note that setting 
$ \alpha=1, \beta=0 $ in equations (\ref{oL})-(\ref{muR}) leads to agreement between (\ref{PabsSL2Z}) and (\ref{Pabs2CFTSL2Z}) in $J$ picture. Moreover, setting 
$ \alpha=0, \beta=1 $ in equations (\ref{oL})-(\ref{muR}) leads to agreement between (\ref{PabsSL2Z}) and (\ref{Pabs2CFTSL2Z}) in $Q$ picture.

\section{Scattering cross-section on extremal background}\label{extscat}
In previous section, we found more evidence to support the correspondence between the non-extremal rotating black holes (\ref{KNNUTAdSmetric}) and a two-dimensional  CFT, in different conformal pictures. In this section, we consider the extremal black holes and provide more evidence in support of the correspondence between the extremal rotating black holes (\ref{KNNUTAdSmetric}) where $ r_+ =r_* $, and the dual CFT.  
The radial equation (\ref{radeq}) in the extremal limit, can be written as
\begin{eqnarray}
\left[\partial_r\left(\Delta_r \partial_r\right) +\f{A_e}{r-r_+}+\f{B_e^2}{(r-r_+)^2}-l'(l'+1)\right]R(r) =0,\label{radialequationext}
\end{eqnarray}
where
\begin{eqnarray}
&& A_e = \frac{2(2\omega r_+ -qe)[(r_+^2+(a+n)^2)\omega - am -qer_+]\Xi^2}{K^2}, \\
&& B_e = \frac{[(r_+^2+(a+n)^2)\omega - am -qer_+]\Xi}{K}. \
\end{eqnarray}
We change the coordinate $r$ to $z$, given by $ z =\frac{-2iB_e}{r-r_+} $. The radial equation (\ref{radialequationext}) becomes
\begin{eqnarray}
\frac{d^2R(z)}{dz^2} + \left[\frac{\frac{1}{4}-m_s^2}{z^2} + \frac{\bar{k}}{2z}-\frac{1}{4}\right]R(z) =0
,\label{radialequationext1}
\end{eqnarray}
where
\begin{eqnarray}
\bar{k}=\frac{i(2r_+\omega-qe)\Xi}{K},~~~m_s =\sqrt{l'(l'+1) + \frac{1}{4}}. \
\end{eqnarray}
We find the solutions to (\ref{radialequationext1}) as
\begin{equation}
R(z)= D_1 R_+(z)+ D_2 R_-(z), 
\end{equation}
where
\begin{equation}
R_\pm (z) = e^{-z/2}z^{m_s\pm 1/2}~_1F_1\left(m_s\pm 1/2-\bar k, 2m_s\pm 1;z \right). \label{sol1}
\end{equation}
In (\ref{sol1}),  $_1F_1(a,b;z) $ is the Kummer’s function. In the near-horizon region $ r\rightarrow r_+ $, where $ z \rightarrow \infty $, the Kummer’s function could be expanded as
\begin{eqnarray}
_1F_1(a,b;z)\sim \frac{\Gamma(b)}{\Gamma(a-b)}z^{-a}e^{-ia\pi}+\frac{\Gamma(b)}{\Gamma(a)}z^{a-b}e^{z}.\
\end{eqnarray}

On the other hand, for asymptotic region where $ r \rightarrow \infty$ (or $ z\rightarrow 0 $), the Kummer’s function $ _1F_1 (a,b;0)\rightarrow 1 $.  To avoid any outgoing waves at $ r=r_+ $, we should choose the constants $D_1$ and $D_2$ as
\begin{eqnarray}
D_1 = -\frac{\Gamma(1-2m_s)}{\Gamma(1/2-m_s-\bar k)}, ~~~ D_2 = \frac{\Gamma(1+2m_s)}{\Gamma(1/2+m_s-\bar k)}. \
\end{eqnarray}
So, in the asymptotically infinity region, the solutions (\ref{sol1}) become
\begin{eqnarray}
R(r)\sim D_1r^{-h}+D_2 r^{1-h},
\end{eqnarray}
where  $ h $ is the conformal weight 
\begin{equation}
h = m_s+\frac{1}{2}.
\end{equation}

The retarded Green function is equal to \cite{ChenChu,ChenLong}
\begin{eqnarray}
G_R \sim \frac{D_{\textcolor{black}{2}}}{D_{\textcolor{black}{1}}} \propto \frac{\Gamma(1-2h)}{\Gamma(2h-1)} \frac{\Gamma(h-\bar{k})}{\Gamma(1-h-\bar{k})}.
\end{eqnarray}
Hence, we can find the scattering cross-section, using $P_{abs}  \sim \Im(G_R)$. To find agreement between the CFT scattering cross-section (\ref{Pabs2CFTSL2Z}) in the general picture and $P_{abs}$, we choose 
$ \omega_L,\mu_L$ and $q_L $ by
\begin{eqnarray}
&&\omega _{L} = \frac{2r_+[r_+^2 +(a+n)^2]}{2a\alpha r_++e\beta (r_+^2 -(a+n)^2)}\omega, \label{oLext}\\
&&\mu _{L} = \frac{e[r_+^2 +(a+n)^2]}{[2a\alpha r_++e\beta (r_+^2 -(a+n)^2)]\Xi},~~~q_L = q. \label{muLext} \
\end{eqnarray}
Moreover, the left frequency is given by $ \tilde \omega _{L} = \omega _{L} - q_{L} \mu _{L} $, and  
$ h_{L} = h $. 
We also note that setting 
$ \alpha=1, \beta=0 $ in equations (\ref{oLext}) and (\ref{muLext}) leads to agreement between CFT scattering cross-section and $P_{abs}$ 
in $J$ picture. Moreover, setting 
$ \alpha=0, \beta=1 $ in equations (\ref{oLext}) and (\ref{muLext}) leads to agreement between CFT scattering cross-section and $P_{abs}$ 
in $Q$ picture.

\section{Conclusions}\label{conc}
We explicitly construct three different conformal pictures for the rotating charged Kerr-Newman-NUT-AdS black holes, that extend the holographic duality between the most complete class of four-dimensional non-extremal (as well as extremal) charged rotating black holes with the NUT twist and the cosmological constant, and conformal field theories. To establish the holography, we consider a charged scalar field in the background of the black holes and construct the scalar wave equation from the generators of the conformal symmetry. 
We find that the charged scalar field in the background of the black holes, reveals the existence of three different CFTs, in which each CFT is dual to each black hole hairs, such as angular momentum and the electric charge. 

To further support the holograph, we find the microscopic entropy and the absorption cross section of the charged scalar fields, in the background of Kerr-Newman-NUT-AdS black holes in three different pictures.  In all three pictures, we find perfect agreement between the gravitational entropy and the absorption cross section and the corresponding results from a two-dimensional CFT.  A very interesting project, is to find the extended family of the conformal symmetry for the Kerr-Newman-NUT-AdS black holes, in which the symmetry is deformed by a deformation parameter \cite{last}. The other project is to establish the existence of a super conformal field theory, with a global $U(1)$ symmetry, in which the $J$ and $Q$ pictures are related to the spectral flow transformations of the super conformal field theory. We leave these projects for future research works.

\vspace*{5mm}

\noindent {\large{\bf Acknowledgments}}

A. M. Ghezelbash would like to acknowledge the support, by the Natural Sciences and Engineering Research Council of Canada. Part of this work by M. F. A. R. S., A. S, and F. P. Z. is supported by Riset PMDSU 2018 and PKPI Scholarship from Ministry of Research, Technology, and Higher Education of
the Republic of Indonesia.  F. P. Z. also acknowledges the support from P3MI 2020.\newline


\vspace*{1mm}


\begin{thebibliography}{99}


\bibitem{stro} 
M.~Guica, T.~Hartman, W.~Song and A.~Strominger,
  %``The Kerr/CFT Correspondence,''
  \textit{Phys.\ Rev.} {\bf D 80}, 124008 (2009).

\bibitem{KerrCFT1}
Y.~Matsuo, T.~Tsukioka and C.~-M.~Yoo,
  %``Notes on the Hidden Conformal Symmetry in the Near Horizon Geometry of the Kerr Black Hole,''
\textit{Nucl.\ Phys.} {\bf B 844}, 146 (2011).
  
\bibitem{KerrCFT1A}
B.~C.~da Cunha and A.~R.~de Queiroz,
  %``Kerr-CFT From Black-Hole Thermodynamics,''
  \textit{JHEP} {\bf 1008}, 076 (2010);  
%  [arXiv:1006.0157 [hep-th]]; 
J.~Rasmussen,
  %``On the CFT duals for near-extremal black holes,''
\textit{Mod.\ Phys.\ Lett.} {\bf A 26}, 1601 (2011).
  
\bibitem{KerrCFT1B}
%  [arXiv:1005.2255 [hep-th]]; 
A. M. Ghezelbash,
  %``Hidden conformal symmetry of rotating charged black holes,''
\textit{Mod.\ Phys.\ Lett.} {\bf A 27}, 1250046 (2012);
%  [arXiv:1005.1404 [gr-qc]]; 
R.~Li, M.~-F.~Li and J.~-R.~Ren,
  %``Entropy of Kaluza-Klein Black Hole from Kerr/CFT Correspondence,''
  \textit{Phys.\ Lett.} {\bf B 691}, 249 (2010).
  
\bibitem{KerrCFT1C}
J.~Rasmussen,
  %``A near-NHEK/CFT correspondence,''
\textit{Int.\ J.\ Mod.\ Phys.} {\bf A 25}, 5517 (2010);
%  [arXiv:1004.4773 [hep-th]]; 
C.~Krishnan,
  %``Hidden Conformal Symmetries of Five-Dimensional Black Holes,''
\textit{JHEP} {\bf 1007}, 039 (2010);
%  [arXiv:1004.3537 [hep-th]]; 
M.~Becker, S.~Cremonini and W.~Schulgin,
  %``Extremal Three-point Correlators in Kerr/CFT,''
\textit{JHEP} {\bf 1102}, 007 (2011).
%  [arXiv:1004.1174 [hep-th]]; 

\bibitem{KerrCFT2}
J.~-J.~Peng and S.~-Q.~Wu,
  %``Extremal Kerr/CFT correspondence of five-dimensional rotating (charged) black holes with squashed horizons,''
  \textit{Nucl.\ Phys.} {\bf B 828}, 273 (2010);
%  [arXiv:0911.5070 [hep-th]]; 
T.~Hartman, W.~Song and A.~Strominger,
  %``Holographic Derivation of Kerr-Newman Scattering Amplitudes for General Charge and Spin,''
  \textit{JHEP} {\bf 1003}, 118 (2010).
  
\bibitem{KerrCFT1D}
%  [arXiv:0908.3909 [hep-th]]; 
J.~Rasmussen,
  %``Isometry-preserving boundary conditions in the Kerr/CFT correspondence,''
  \textit{Int.\ J.\ Mod.\ Phys.} {\bf A 25}, 1597 (2010);
%  [arXiv:0908.0184 [hep-th]]; 
  Y.~Matsuo, T.~Tsukioka and C.~-M.~Yoo,
  %``Yet Another Realization of Kerr/CFT Correspondence,''
  \textit{Europhys.\ Lett.}\  {\bf 89}, 60001 (2010).
  
\bibitem{KerrCFT1R}
%  [arXiv:0907.4272 [hep-th]]; 
  I.~Bredberg, T.~Hartman, W.~Song and A.~Strominger,
  %``Black Hole Superradiance From Kerr/CFT,''
  \textit{JHEP} {\bf 1004}, 019 (2010);
%  [arXiv:0907.3477 [hep-th]]; 
  O.~J.~C.~Dias, H.~S.~Reall and J.~E.~Santos,
  %``Kerr-CFT and gravitational perturbations,''
  \textit{JHEP} {\bf 0908}, 101 (2009).
  
\bibitem{KerrCFT1Y}
%  [arXiv:0906.2380 [hep-th]]; 
  L.~-M.~Cao, Y.~Matsuo, T.~Tsukioka and C.~-M.~Yoo,
  %``Conformal Symmetry for Rotating D-branes,''
  \textit{Phys.\ Lett.} {\bf B 679}, 390 (2009);
%  [arXiv:0906.2267 [hep-th]];  
  C.~Krishnan and S.~Kuperstein,
  %``A Comment on Kerr-CFT and Wald Entropy,''
  \textit{Phys.\ Lett.} {\bf B 677}, 326 (2009).
  
\bibitem{KerrCFT1AO}
 B. Chen, A.M. Ghezelbash, V. Kamali and M.R. Setare,
\textit{Nucl. Phys.} {\bf  B 848}, 108 ( 2011); 
%  [arXiv:1004.3963 [hep-th]]; 
  J.~Mei,
  %``The Entropy for General Extremal Black Holes,''
  \textit{JHEP} {\bf 1004}, 005 (2010).
%  [arXiv:1002.1349 [hep-th]]; 
  
 \bibitem{KerrCFT1AOY}  
 E.~Barnes, D.~Vaman and C.~Wu,
  %``All 4-dimensional static, spherically symmetric, 2-charge abelian Kaluza-Klein black holes and their CFT duals,''
  \textit{Class.\ Quant.\ Grav.}\  {\bf 27}, 095019 (2010);
%  [arXiv:0908.2425 [hep-th]]; 
  X.~-N.~Wu and Y.~Tian,
  %``Extremal Isolated Horizon/CFT Correspondence,''
  \textit{Phys.\ Rev.} {\bf D 80}, 024014 (2009).
  
 \bibitem{KerrCFT1AOTT}
%  [arXiv:0904.1554 [hep-th]]; 
  T.~Azeyanagi, G.~Compere, N.~Ogawa, Y.~Tachikawa and S.~Terashima,
  %``Higher-Derivative Corrections to the Asymptotic Virasoro Symmetry of 4d Extremal Black Holes,''
  \textit{Prog.\ Theor.\ Phys.}\  {\bf 122}, 355 (2009);
%  [arXiv:0903.4176 [hep-th]]; 
%  [arXiv:0902.4387 [hep-th]; 
  D.~Astefanesei and Y.~K.~Srivastava,
  %``CFT Duals for Attractor Horizons,''
  \textit{Nucl.\ Phys.} {\bf B 822}, 283 (2009).
  
 \bibitem{KerrCFT1AOW}
%  [arXiv:0902.4033 [hep-th]]; 
  K.~Hotta,
  %``Holographic RG flow dual to attractor flow in extremal black holes,''
  \textit{Phys.\ Rev.} {\bf D 79}, 104018 (2009);
%  [arXiv:0902.3529 [hep-th]]; 
  G.~Compere, K.~Murata and T.~Nishioka,
  %``Central Charges in Extreme Black Hole/CFT Correspondence,''
  \textit{JHEP} {\bf 0905}, 077 (2009); A. M. Ghezelbash and H. M. Siahaan,  \textit{Phys. Rev.} {\bf D 89},  024017  (2014).
%  [arXiv:0902.1001 [hep-th]].

\bibitem{KerrCFT3}
  C.~-M.~Chen, Y.~-M.~Huang and S.~-J.~Zou,
  %``Holographic Duals of Near-extremal Reissner-Nordstrom Black Holes,''
  \textit{JHEP} {\bf 1003}, 123 (2010); A. M. Ghezelbash, \textit{Gen. Rel. Grav.} {\bf 48}, 102 (2016).
%  [arXiv:1001.2833 [hep-th]]; 

\bibitem{canisius}
B. Canisius and A. M. Ghezelbash,  \textit{Phys. Rev.} {\bf D 101},  024020  (2020).

\bibitem{SaktiAnnPhys2020}
M.~F.~A.~R.~Sakti, A.~Suroso and F.~P.~Zen,
  %``Kerr-Newman-NUT-Kiselev black holes in Rastall Theory of gravity and Kerr/CFT Correspondence,''
    \textit{Ann. Phys. } {\bf 413}, 168062 (2020).
 
\bibitem{cms}
A.~Castro, A.~Maloney and A.~Strominger, \textit{Phys. Rev.} \textbf {D 82}, 024008 (2010).

\bibitem{othe}
K. -N.~Shao and Z.~Zhang, \textit{Phys. Rev.} \textbf{D 83}, 106008 (2011); 
B.~Chen, J.~Long and J. -J.~Zhang, \textit{Phys. Rev.} \textbf{D 82}, 104017 (2010).

\bibitem{KerrCFT1AOQQ}
 A.~P.~Porfyriadis and F.~Wilczek, [arXiv:1007.1031].
 
\bibitem{KerrCFT1AQAO}
R.~Monteiro, [arXiv:1006.5358],
Y. -Q. Wang and Y. -X. Liu, 
\textit{JHEP} {\bf 1008}, 087 (2010).

\bibitem{KSChen}
D.~Chen, P.~Wang, H.~Wu and H. Yang,
  %``Hidden conformal symmetry of rotating charged black holes,''
 \textit{Gen. Rel. Grav.} {\bf 43}, 181 (2011).

\bibitem{Chen-JQpic}
  C.~-M.~Chen, Y.~-M.~Huang, J.~-R.~Sun, M.~-F.~Wu and S.~-J.~Zou,
  %``Twofold Hidden Conformal Symmetries of the Kerr-Newman Black Hole,''
  \textit{Phys.\ Rev.} {\bf D 82}, 066004 (2010)

\bibitem{chen-SL2Z}
B.~Chen and J. -J. Zhang, \textit{JHEP} \textbf{1108}, 114 (2011); B.~Chen and J. -J. Zhang,
  %``Novel CFT Duals for Extreme Black Holes,''
  \textit{Nucl.\ Phys.} {\bf B  856}, 449 (2012).

\bibitem{sen}
A.~Sen, \textit{Phys. Rev. Lett.} \textbf{69}, 1006 (1992).

\bibitem{Ghezelbash}
A. M. Ghezelbash and H. M. Siahaan,  \textit{Gen. Rel. Grav.} {\bf30}, 135005 (2013).

\bibitem{GH2}
A.~M.~Ghezelbash,
  %``Kerr/CFT Correspondence in Low Energy Limit of Heterotic String Theory,''
  \textit{JHEP} {\bf 0908}, 045 (2009).
%  [arXiv:0901.1670 [hep-th]].

\bibitem{Demianski1976}
J.~F.~Pleba{\'n}ski and M.~Demia{\'n}ski, \textit{Ann. Phys.} \textbf{98}, 98 (1976). 
%Rotating,Charged, and Uniformly Accelerating Mass in General Relativity

\bibitem{Griffiths2005}
J.~B.~Griffiths and J.~Podolsk{\'y}, \textit{Class. Quantum Grav.} \textbf{22}, 3467 (2005).
%Accelerating and rotating black holes

\bibitem{Podolsky2009}
J.~Podolsk{\'y} and H.~Kadlecov{\'a}, \textit{Class. Quantum Grav.} \textbf{26}, 105007 (2006).
%arXiv:0903.3577 [gr-qc].
%Radiation generated by accelerating and rotating charged black holes in (anti-)de Sitter space

  
\bibitem{Compere}
  G.~Compere,
  %``The Kerr/CFT correspondence and its extensions: a comprehensive review,''
  \textit{Living Rev.\ Rel.}\  {\bf 15}, 11 (2012).
  %[arXiv:1203.3561 [hep-th]].
  
\bibitem{ChenJHEP2010}
B.~Chen and Jiang~Long,
  %``On holographic description of the Kerr-Newman-AdS-dS black holes,''
    \textit{JHEP} {\bf 1008}, 065 (2010).
    
\bibitem{ChenNuc2011-a}
B.~Chen, C-M.~Chen and Bo~Ning,
  %``Holographic Q-picture of Kerr–Newman–AdS–dS black hole,''
    \textit{Nuc. Phys. } {\bf B 853}, 196-209 (2011).
    
\bibitem{ChenNuc2011-b}
B.~Chen, A.~M.~Ghezelbash, V.~Kamali and M.~R.~Setare,
  %``Holographic description of Kerr–Bolt–AdS–dS spacetimes''
    \textit{Nuc. Phys. } {\bf B 848}, 108-120 (2011).

\bibitem{RN-CFT1}  
  C.~-M.~Chen, J.~-R.~Sun and S.~-J.~Zou,
  %``The RN/CFT Correspondence Revisited,''
  \textit{JHEP} {\bf 1001}, 057 (2010);
  %[arXiv:0910.2076 [hep-th]];
\bibitem{RN-CFT2}
  B.~Chen and J.~-J.~Zhang,
  %``RN/CFT Correspondence From Thermodynamics,''
  \textit{JHEP} {\bf 1301}, 155 (2013).
 % [arXiv:1212.1959 [hep-th]].
\bibitem{RN-CFT3}
  M.~F.~A.~R. Sakti, A.~Suroso and F.~P.~Zen,
  %``CFT duals on extremal rotating NUT black holes,''
  \textit{Int. J. Mod. Phys.} \textbf{D 27}, 1850109 (2018).
\bibitem{RN-CFT4}
  M.~F.~A.~R. Sakti, A.~Suroso and F.~P.~Zen,
  %``Kerr/CFT Correspondence on Kerr-Newman-NUT-Quintessence Black Hole,''
  \textit{Eur. Phys. J. Plus} \textbf{134}, 580 (2019).
  
\bibitem{chen-novelsl2z} 
  B.~Chen and J.-J.~Zhang,
  %``Novel CFT Duals for Extreme Black Holes,''
 \textit{ Nucl.\ Phys.}\  {\bf B 856}, 449 (2012).
  %[arXiv:1106.4148 [hep-th]].

\bibitem{Chen-genhidden4d5d} 
  B.~Chen and J.-J.~Zhang,
  %``General Hidden Conformal Symmetry of 4D Kerr-Newman and 5D Kerr Black Holes,''
  \textit{JHEP} {\bf 1108}, 114 (2011).
  %[arXiv:1107.0543 [hep-th]].
  
\bibitem{Ghezelbashdeform}
A.~M.~Ghezelbash and H.~M.~Siahaan,
%``Deformed Hidden Conformal Symmetry for Rotating Black Holes,''
\textit{Gen. Rel. Grav.} \textbf{46}, 1783 (2014).

\bibitem{Malda-Strom} 
  J.~M.~Maldacena and A.~Strominger,
  %``Universal low-energy dynamics for rotating black holes,''
 \textit{ Phys.\ Rev.}\ {\bf D 56}, 4975 (1997).
  %[hep-th/9702015].


\bibitem{ChenChu}
B. Chen and C. S. Chu, 
%Real-time correlators in Kerr/CFT correspondence
\textit{JHEP} {\bf 1005}, 004 (2010).
%  %[arXiv:1001.3208 [hep-th]].

\bibitem{ChenLong}
B. Chen and J.  Long,
%Real-time correlators and hidden conformal symmetry in the Kerr/CFT correspondence
\textit{JHEP} {\bf 1006}, 018 (2010).
%[arXiv:1004.5039 [hep-th]].


\bibitem{last}
M.~F.~A.~R. Sakti, A. M. Ghezelbash, A.~Suroso and F.~P.~Zen
%Deformed conformal symmetry of Kerr-Newman-NUT-AdS black holes
\textit{Gen. Rel. Grav.} {\bf 51}, 151 (2019).




\end{thebibliography}
\end{document}